\documentclass[%
 reprint,
 amsmath,amssymb,
 aps,
floatfix,
]{revtex4-2}
\usepackage{graphicx}% Include figure files
\graphicspath{{Pictures/}} % Specifies the directory where pictures are stored
\usepackage{dcolumn}% Align table columns on decimal point
\usepackage{subfigure}
\usepackage{bm}% bold math
\usepackage[colorlinks=true,
            linkcolor=blue,
            citecolor=blue,
            urlcolor=blue,
            pdfborder={0 0 0}]{hyperref}

\begin{document}

\preprint{APS/123-QED}

\title{Lensing, Shadow and Photon Rings in a Magnetized Black Hole Spacetime}% Force line breaks with \\
%\thanks{A footnote to the article title}%

\author{Muhammad Haider Khan}
\email{muhammadhaider.khan@students.uniroma2.eu}
\email{khan7@uni-bremen.de}
\affiliation{Faculty 1, University of Bremen, 28359 Bremen, Germany}

\author{Volker Perlick}
\email{perlick@uni-bremen.de}
\affiliation{Faculty 1, University of Bremen, 28359 Bremen, Germany}
\date{\today}
\begin{abstract}
{ 
We analyze gravitational lensing, in particular the shadow and photon rings, in the Ernst spacetime, also known as the Schwarzschild-Melvin spacetime, which describes a Schwarzschild black hole immersed in a homogenous magnetic field. Although the geodesic equation in this spacetime is chaotic, there are some relevant features that can be determined analytically. Among other things, we give analytic formulas for the vertical diameter of the shadow for an observer at arbitrary inclination and for the horizontal diameter of the shadow for an observer in the equatorial plane. Moreover, we use the strong-deflection formalism for analytically calculating the so-called photon rings of order $\ge 2$ and we use the recently introduced gap parameter $\Delta _2$ for distinguishing lensing of an Ernst black hole from that of a Schwarzschild black hole.}
\end{abstract}
\maketitle

%\tableofcontents

\section{\label{sec:INTRODUCTION}INTRODUCTION\protect\\}
Gravitational lensing and shadow formation by black holes depend fundamentally on the behaviour of null geodesics. We investigate such geodesics in the { Ernst spacetime, also known as the Schwarzschild-Melvin spacetime,} which describes a Schwarzschild black hole of mass $M$ immersed in a uniform magnetic field of strength $B$ \cite{Ernst1976}. For $M=0$, the spacetime reduces to the Bonnor-Melvin magnetic universe \cite{Bonnor1954,Melvin1964}. 
%The magnetic field enters the metric via a factor $\Lambda(r,\theta,B)$. 

{ Studying the influence of magnetic fields on the lensing features of black holes is of high astrophysical relevance. There is good evidence that magnetic fields are ubiquitous in the universe. Assuming that a (supermassive) black hole is immersed in a magnetic field that is asymptotically homogeneous is certainly idealized, but it is not an unrealistic model. In some cases these magnetic fields could be strong enough to have a measurable influence on the gravitational field. Then the magnetic field has an effect on the lensing properties of the black hole, in particular in the strong-bending regime. }

As the geodesic equation in the Ernst spacetime is chaotic, a complete picture of the lensing features can be gained only numerically. Such a numerical study has been performed by Lima Junior \textit{et al.} \cite{LimaJunior2022}. It is the main purpose of the present paper to complement this earlier work by demonstrating that, in spite of the fact that the geodesic equation is chaotic, a few relevant lensing features can be determined analytically. To that end, we make use of two facts: Firstly, we utilize the fact that the geodesic equation becomes completely integrable if we restrict ourselves to the equatorial plane. In particular, this allows one to determine the circular lightlike geodesics in the equatorial plane analytically. It was shown already by Dhurandhar and Sharma \cite{Dhurandhar1983} and by Esteban \cite{Esteban1984} that the radius coordinates where such geodesics exist are determined by a cubic equation. If the magnetic field strength $B$ lies between zero and a critical value $B_c$, this equation has two real positive roots outside the horizon: there is an inner unstable photon orbit at a radius $r_2$ (with $r_2 \to 3M$ if $B \to 0$) and an outer stable photon orbit $r_1$ (with $r_1 \to \infty$ if $B \to 0$). At the critical field strength $B = B_c$ these two roots merge, and for $B\ge B_c$ no circular photon orbit exists outside the horizon. Secondly, we utilize the fact that the equation for \emph{lightlike} geodesics becomes completely integrable also in every meridional plane. (A meridional plane is a plane that consists of two half-planes $\phi = \phi _0$ and $\phi = \phi _0 + \pi$.) This follows immediately from the observation, already emphasized by Lima Junior \textit{et al.} \cite{LimaJunior2022}, that in each meridional plane the Ernst metric is conformally equivalent to the Schwarzschild metric. Hence, null geodesics in a meridional plane follow the same paths as in Schwarzschild; only the affine parametrization changes. In particular, each meridional plane contains an unstable circular photon orbit at $r=3M$. 

In contrast to, e.g., the Kerr spacetime, in the Ernst spacetime, there are no non-circular spherical lightlike geodesics, i.e., no lightlike geodesics that stay on a sphere $r = \mathrm{constant}$ without being circular. Below, we will give a formal proof of this non-existence result, which is, of course, related to the fact that in the Ernst spacetime { the geodesic equation does not admit sufficiently many constants of motion for complete integrability. By contrast, the Kerr spacetime admits the so-called Carter constant which makes the geodesic equation completely integrable. The existence of the Carter constant is related to the existence of spherical lightlike geodesics which are crucial} for analytically calculating relevant lensing features in the strong-deflection regime, in particular the shadow. For reviews of how to calculate the shadow of a black hole analytically we refer to Cunha and Herdeiro \cite{CunhaHerdeiro2018} and to Perlick and Tsupko \cite{PerlickTsupko2022}. 

In the Ernst spacetime, { it is at least possible to determine some of the lightlike geodesics analytically, namely the ones in the equatorial plane and the ones that are contained in a meridional plane.} We will demonstrate below that this allows us to calculate at least some lensing features analytically: We will show that it allows us to calculate the vertical and, for observers in the equatorial plane, also the horizontal radius of the shadow. We will also demonstrate that it will enable us to calculate for a polar observer the so-called \emph{photon rings}. 

{ 
%Actually, the notion of photon rings is associated with a certain confusion of terminology and it might be worthwhile to add a few clarifying words on this issue. First of all, ``photon rings'' have to be distinguished from ``light rings''. The latter term is used, in particular by the Portuguese school, as synonymous to ``circular light rays''. So light rings are circles in 3-dimensional space. By contrast, photon rings are circles on the observer's sky. 
The notion of photon rings is based on the observation that in the spacetime of a black hole -- Schwarzschild is the standard example -- every observer sees infinitely many images of any light source if both are located outside the horizon and not exactly opposite to each other, see e.g. Perlick \cite{Perlick2004}. In a spherically symmetric and static spacetime, the images can be labelled by a non-negative integer, called the ``order'', which counts how often the light ray associated with this image meets the optical axis, i.e., the straight line through the position of the observer and the center of the coordinate system. The image of order 0 is known as the primary image, the image of order 1 as the secondary image, and all other images as higher-order images. For the Schwarzschild spacetime, this observation can be traced back to the early days of relativity. It was later discussed in great detail, e.g., by Darwin \cite{Darwin1959} and by Ohanian \cite{Ohanian1986}. An important contribution was made by Virbhadra and Ellis \cite{VirbhadraEllis2000} who cast Schwarzschild lensing in the form of a lens equation. They termed the images of order $n \ge 2$ ``relativistic images''; other authors call them ``higher-order images''. The notion of photon rings comes into the game if we consider a continuous distribution of light sources, e.g. a luminous accretion disk. In the spacetime of a spherically symmetric and static black hole, each point of such a disk produces an image of order $n$, for each non-negative integer $n$. These images arrange themselves into an infinite sequence of nested rings on the observer's sky which converge upon the shadow of the black hole.  With increasing $n$, these rings become thinner and fainter. Until now, it was not possible to isolate these rings observationally, but there is some hope that in the not-too-far future it might be possible to isolate the outermost ones. This gives us a good motivation for calculating the photon rings in the Ernst spacetime. We use the plural, photon rings, because we consider the infinitely many rings labelled by the order $n$. By contrast, other authors use the term photon ring (singular) for a compound of these infinitely many rings from some order $n$ onwards, see in particular Gralla et al. \cite{GrallaHolzWald2019}, who introduced this term, and Johnson et al. \cite{JohnsonEtAl2020} who thoroughly investigated the perspectives of actually observing photon rings. In this paper we will make use of the fact that the notion of photon rings is well-defined in the Ernst spacetime,} although the latter is not spherically symmetric, if we consider a luminous ring in the equatorial plane and a polar observer. This allows us to analytically calculate the photon rings with the help of the strong-deflection formalism and to compare with the Schwarzschild case.  

This paper is organized as follows: In Sec. \ref{sec:Spacetime}, we review the Ernst metric. In Sec. \ref{sec:Geodesics in the Ernst Spacetime}, { we collect some results on lightlike and timelike geodesics in the Ernst spacetime that will be used later on. In particular, we prove that in an Ernst spacetime with $B \neq 0$ no non-circular spherical lightlike geodesics exist and that the only geodesics that can reach infinity are the meridional ones. As briefly outlined in Sec. \ref{sec:Angular Deflection}, the latter observation implies that the usual definition of a deflection angle makes sense only for meridional lightlike geodesics, for which the formula for the deflection angle is identical with the Schwarzschild one. In Sec. \ref{sec:Shadow Angular Radii}, we derive analytical formulas for the vertical radius of the shadow for an observer at arbitrary inclination and for the horizontal radius of the shadow for an observer in the equatorial plane. Sec. \ref{sec:Photon Rings in Ernst Spacetime} is devoted to the calculation of the photon rings as seen by a polar observer. We conclude with Sec. \ref{sec:Conclusion}.
}
%-----------------------------------------------------------------------------------
\section{\label{sec:Spacetime}ERNST SPACETIME}
The Ernst metric \cite{Ernst1976}, also known as the Schwarzschild-Melvin metric, is a static, axisymmetric solution of the Einstein-Maxwell equations. It describes a Schwarzschild black hole of mass $M$ immersed in an external uniform magnetic field $B$. In Schwarzschild coordinates $(t,r,\theta,\varphi)$ it reads
\begin{eqnarray}
    ds^2=-\Lambda^2\left(1-\frac{2M}{r}\right)d t^2+\frac{\Lambda^2}{\left(1-\frac{2M}{r}\right)}d r^2 \nonumber\\
    + r^2 \Lambda^2d \theta^2+\frac{r^2\sin^2\theta}{\Lambda^2}d\phi^2
    \label{eq:Ernst}
\end{eqnarray}
where the factor $\Lambda$ is defined as
\begin{eqnarray}
    \Lambda=1+\frac{1}{4}r^2B^2\sin^2\theta.
\end{eqnarray}
The event horizon remains at $r=2M$, as in Schwarzschild. However, unlike Schwarzschild, this spacetime is not asymptotically flat: at large $r$ the metric approaches the { Bonnor-Melvin} magnetic universe \cite{Bonnor1954,Melvin1964}. In the limit $B\to0$, we recover the Schwarzschild metric.
%--------------------------------------------------------------------------------------------------------
\section{\label{sec:Geodesics in the Ernst Spacetime}GEODESICS IN THE ERNST SPACETIME}
The { Lagrangian $\mathcal{L} = \frac{1}{2} g_{\mu \nu} \dot{x}{}^{\mu} \dot{x}{}^{\nu}$ governs the geodesics. There are two conserved quantities, $E$ and $L$,} corresponding to the time-translation and axial symmetries, respectively,
\begin{eqnarray}
    -\frac{\partial\mathcal{L}}{\partial \dot{t}}=E=\Lambda^2\left(1-\frac{2M}{r}\right)\dot{t},\\
    \frac{\partial\mathcal{L}}{\partial \dot{\phi}}=L=\frac{r^2\sin^2\theta}{\Lambda^2}\dot\phi.
    \label{eq:L}
\end{eqnarray}
{ The Lagrangian is a third constant of motion,}

\begin{eqnarray*}
    \mathcal{L}=\frac{1}{2}\bigg[-\frac{1}{\Lambda^2(1-\frac{2M}{r})}E^2+\frac{\Lambda^2}{r^2\sin^2\theta}L^2
    \end{eqnarray*}
    \begin{eqnarray}
    +\frac{\Lambda^2}{(1-\frac{2M}{r})} \dot r^2+r^2\Lambda^2\dot \theta^2\bigg]=\frac{1}{2}\epsilon
    \label{eq:Lagrangian}
\end{eqnarray}
The constant $\epsilon$ has values of { $0$, $-1$ and $+1$  for lightlike, timelike and spacelike geodesics}, respectively.
{ If $B \neq 0$ there is no fourth constant of motion, so the geodesic equation is not completely integrable.}

After rewriting  Eq. (\ref{eq:Lagrangian}) as
\begin{eqnarray}
\dfrac{r^2 \mathrm{sin} ^2 \theta \, E^2}{\Big( 1 + \dfrac{r^2}{4} B^2 \mathrm{sin}^2 \theta \Big)^4 \Big( 1 - \dfrac{2M}{r} \Big)} +
\dfrac{\epsilon \, r^2 \mathrm{sin}^2 \theta}{\Big( 1 + \dfrac{r^2}{4} B^2 \mathrm{sin}^2 \theta \Big)^2} 
\nonumber \\
=
L^2 + \dfrac{r^2 \mathrm{sin}^2 \theta \, \dot{r}{}^2}{\left(1 - \dfrac{2M}{r}\right)} 
+ r^4 \mathrm{sin}^2 \theta \, \dot{\theta}{}^2 \, , \qquad \quad
\end{eqnarray}
we see immediately that geodesics with $L \neq 0$ cannot reach infinity in an Ernst spacetime with $B \neq 0$: If $L \neq 0$, the right-hand side is bounded away from 0 on the interval $2M<r<\infty$, so the left-hand side must be bounded away from 0 as well. However, along a geodesic with $L \neq 0$, Eq. (\ref{eq:L}) requires $\mathrm{sin} \, \theta \neq 0$; as a consequence, if such a geodesic reaches infinity, the left-hand side would approach 0 if $B \neq 0$. This generalizes a result of Dhurandhar and Sharma~\cite{Dhurandhar1983}, who had shown that lightlike geodesics in the equatorial plane with $L \neq 0$ cannot reach infinity.

%----------------------------------------------------------------------------------------------------------
\subsection{\label{subsec: Circular Lightlike Geodesics}Circular lightlike geodesics about the $z$-axis}
\begin{figure}[htbp]
    \centering
    \includegraphics[width=0.5\textwidth]{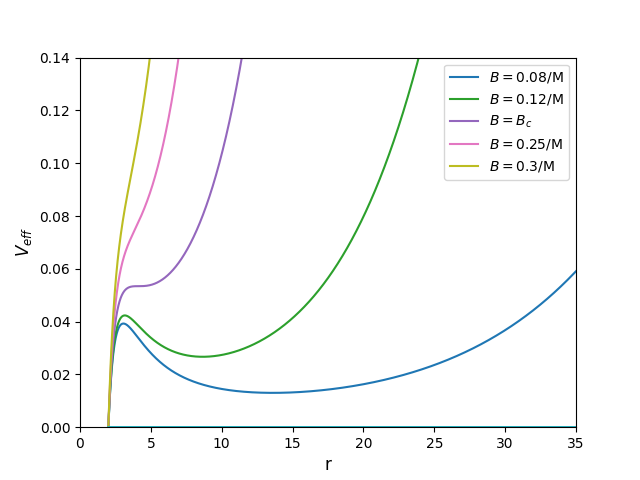}
    \caption{\label{fig:Effective_potential}Equatorial effective potential $V_\mathrm{eff}$ from Eq. (\ref{eq:Effective_potential}) vs. $r$ for various $B$ values. Here we give $r$ in units of $M$ and $V_\mathrm{eff}$ in units of $1/M^2$.}
\end{figure}

Circular lightlike orbits about the $z$-axis require
\begin{gather*}
    \epsilon=0,\\
    \dot r=\ddot r=0,\\
{
    \dot \theta=\ddot \theta=0.
}
\end{gather*}
Applying these conditions, we find that they hold only in the equatorial plane
$(\theta=\pi/2)$, leading to the cubic equation
\begin{equation}
3B^2 r^3 - 5MB^2 r^2 - 4r +12M = 0.
\label{eq:cubic_equation}
\end{equation}
Dhurandhar and Sharma \cite{Dhurandhar1983} and Esteban \cite{Esteban1984} showed that this cubic equation admits two real solutions $r_1$ and $r_2$ with $r_1>r_2>2M$ if
$0 < B < B_c$, where
\begin{eqnarray}
    B_c \approx  \frac{0.189366386}{M} \, ,
    \label{eq: Critical_B}
\end{eqnarray}
see Appendix \ref{app: Solutions of the Cubic Equation} for the closed-form expressions of these solutions. 
If $B$ increases from $0$ to $B_c$, $r_1$ decreases from infinity to a limiting value
$r_c$ while $r_1$ increases from $3 M$ to $r_c$, where
\begin{equation}
r_c \approx 4.119632981 \, M \, .
\label{eq:rc}
\end{equation}
For $B = B_c$ the two circular photon orbits merge at $r = r_c$ and for $B > B_c$ no circular photon orbits exist outside the horizon. Dhurandhar and Sharma \cite{Dhurandhar1983} define an effective potential for light rays in the equatorial plane
\begin{eqnarray}
    V_\mathrm{eff}=\frac{\Lambda^4}{r^2}\left(1-\frac{2M}{r}\right) \, .
\label{eq:Effective_potential}
\end{eqnarray}
{ Then the orbit equation takes the following form:}
\begin{eqnarray}
    \left(\frac{dr}{d\phi}\right)^2=\frac{r^4}{\Lambda^8}\left(\frac{E^2}{L^2}-V_\mathrm{eff}\right).
\end{eqnarray}
$V_\mathrm{eff}$ is plotted in Fig. \ref{fig:Effective_potential} for various $B$ values. 
The maximum and the minimum of the plots correspond, respectively, to the unstable ($r_2$) and stable ($r_1$) circular lightlike geodesics in the equatorial plane. At $B=B_c$ these roots coincide and disappear. For $B>B_c$ no critical points exist, and hence we do not have any circular lightlike geodesics.
%------------------------------------------------------------------------------------------------------
\subsection{\label{subsec:Meridional Conformality}{ Lightlike geodesics in a meridional plane}}
{
We now consider a meridional plane, i.e., a plane that is the union of two half-planes $\phi = \phi _0$ and $\phi = \phi _0 + \pi$. On such a plane, the coordinate $\theta$ is double-valued: On a circle around the center it goes from 0 to $\pi$ and then back from $\pi$ to 0. Therefore, we introduce, instead of $\theta$, in each meridional plane an azimuthal coordinate $\tilde{\phi}$ that takes values in $\mathbb{R}$ modulo $2 \pi$, quite analogous to our original azimuthal coordinate $\phi$ in the equatorial plane. Then the Ernst metric reduces on each meridional plane to the following form:
\begin{eqnarray}
    ds^2= \Big( 1 + \dfrac{1}{4} B^2 r^2 \mathrm{cos} ^2 \tilde{\phi} \Big) ^2 \times \qquad 
    \nonumber \\
    \Bigg(  - \left(1-\frac{2M}{r}\right)d t^2 
    + \frac{dr^2}{\left(1-\frac{2M}{r}\right)} + r^2 d \tilde{\phi} {}^2 \Bigg)\, .
\end{eqnarray}
We see that on each meridional plane the Ernst metric is conformally equivalent to the Schwarzschild metric, as was already observed by Lima Junior et al. \cite{LimaJunior2022}. As lightlike geodesics are invariant, up to affine parametrization,  under a conformal transformation, we can calculate the lightlike geodesics in a meridional plane from the Schwarzschild Lagrangian,
    \begin{eqnarray*}
    \tilde{\mathcal{L}}=\frac{1}{2}\bigg[-\left(1-\frac{2M}{r}\right)\dot{t}{}^2+\frac{1}{\left(1-\frac{2M}{r}\right)}\dot{r}{}^2+r^2\dot{\tilde \phi}{}^2\bigg].
    \label{eq:tile_lagrangian}
\end{eqnarray*}
The conserved quantities are now 
\begin{eqnarray}
    -    \frac{\partial{\tilde{\mathcal{L}}}}{\partial \dot{t}}= E =\left(1-\frac{2M}{r}\right)\dot{t},
    \label{eq:E_def}
    \end{eqnarray}
    \begin{eqnarray}
    \frac{\partial{\tilde{\mathcal{L}}}}{\partial \dot{\tilde \phi}}=\tilde L=r^2\dot{\tilde{\phi}} .
    \label{eq:Q_def}
\end{eqnarray}
Here $\tilde L$ is the angular momentum arising from the cyclicity of the $\tilde{\phi}$ coordinate in the meridional plane of the { Schwarzschild} spacetime. Now, we have $3$ conserved quantities: $E, \tilde L \text{ and } \tilde{\mathcal{L}}=0$ and a $(2+1)$-dimensional system in the meridional plane, making the lightlike geodesic equations fully integrable.
}
{ Using a standard textbook result from the Schwarzschild metric, there is a circular lightlike geodesic in each meridional plane at $r = 3M$. Together, these lightlike  geodesics sweep out the 2-dimensional surfaces at $r=3M$, $t = \mathrm{constant}$ which, in the Ernst spacetime with the original Schwarzschild-like coordinates, carry the metric 
\begin{eqnarray}
    ds^2|_{t=cte.,r=3M}=
    9 M^2 \Big( 1 + \dfrac{9M^2}{4} B^2 \mathrm{sin} ^2 \theta \Big)^2  d\theta^2
    \nonumber \\
    +\frac{9M^2 \sin^2\theta \, d\phi^2}{\Big( 1+ \dfrac{9M^2}{4} B^2 \mathrm{sin} ^2 \theta \Big) ^2} \, ,
    \qquad \quad 
\end{eqnarray}
see  Lima Junior et al. \cite{LimaJunior2022}. When we refer to this surface as to a ``sphere', we have to be aware of the fact that  it is a coordinate sphere but not a ``round sphere'', i.e., that the induced metric is not the standard metric on a sphere in Euclidean 3-space. Correspondingly, the ``circular geodesics'' that go through the poles are circles only in the coordinate picture, while they are elongated when measured with the intrinsic geometry of the spacetime. Nonetheless, we will call them ``circular'' in the following. 
}
%-----------------------------------------------------------------------------------------------
\subsection{\label{subsec: Spherical Lightlike Geodesics}Spherical lightlike geodesics}
Geodesics that stay at a constant radius value are called spherical geodesics. We prove analytically that the Ernst spacetime does not admit spherical lightlike geodesics { other than the circular lightlike geodesics we have already found}. Spherical geodesics require the condition
\begin{eqnarray}
    \dot r=0, \quad \ddot r=0.
\end{eqnarray}
Using these and $\epsilon=0$ in the Lagrangian (\ref{eq:Lagrangian}), we get the following condition on the constants of motion:
\begin{eqnarray}
    \frac{L^2}{E^2}=\frac{r(r-3M)}{\Lambda^3B^2(r-2M)^2}.
\end{eqnarray}
Note that this equation holds true when $r=3M$ leading to $L=0$ which is the case of the meridional photon orbits at $r=3M$ as discussed in Sec. \ref{subsec:Meridional Conformality}. Except for the meridional plane, since $L$ and $E$ are constants, it demands $\Lambda$ to be a constant as well, meaning that if { $B \neq 0$ the $\theta$ coordinate must be constant along the geodesic, i.e., the geodesic is circular about the $z$-axis. We know already from Sec. \ref{subsec: Circular Lightlike Geodesics} that this is possible only in the equatorial plane, where it gives us two circular lightlike geodesics if $0 < B < B_c$. This completes the proof that no non-circular spherical lightlike geodesics exist.}
%-------------------------------------------------------------------------------
\subsection{\label{subsec: Circular Timelike Geodesics}Circular timelike geodesics { about the $z$-axis}}
{ For our discussion of photon rings below we will also need to know some properties of timelike circular geodesics in the equatorial plane of the Ernst spacetime which we briefly rederive in this section, cf. Esteban \cite{Esteban1984}.}

For timelike geodesics, i.e., orbits of massive uncharged particles, we set $\epsilon=-1$ in Eq. (\ref{eq:Lagrangian}) yielding
\begin{eqnarray}
  -\frac{1}{\Lambda^2(1-\frac{2M}{r})}E^2+\frac{\Lambda^2}{r^2\sin^2\theta}L^2\\\nonumber
    +\frac{\Lambda^2}{(1-\frac{2M}{r})}\dot r^2+r^2\Lambda^2\dot \theta^2=-1,
\end{eqnarray}
For circular geodesics { about the $z$-axis}, we require $\dot r=\ddot r=0$ and {$\dot \theta=\ddot \theta=0$. Solutions exist only in the equatorial plane, $\theta = \pi /2$, where we get the orbit equation}
\begin{gather*}
\frac{1}{2}\left(\frac{dr}{d\phi}\right)^2+V_{E,L}=0
\end{gather*}
Here, we defined an effective potential $V_{E,L}$ given by
\begin{equation}
    -2V_{E,L}=\frac{r^4}{L^2\Lambda^4}\left[\frac{E^2}{\Lambda^4}-\left(1-\frac{2M}{r}\right)\left(\frac{L^2}{r^2}+\frac{1}{\Lambda^2}\right)\right].
    \label{eq: Timelike_Effective_Potential}
\end{equation}
For circular orbits, we have the conditions
\begin{gather*}
    V_{E,L}=0,\quad V_{E,L}'=0.
\end{gather*}
Fig. \ref{fig:V(E,L)} provides a plot of $V_{E,L}$ for different values of $B$.
\begin{figure}[htbp]
    \centering
    \includegraphics[width=0.5\textwidth]{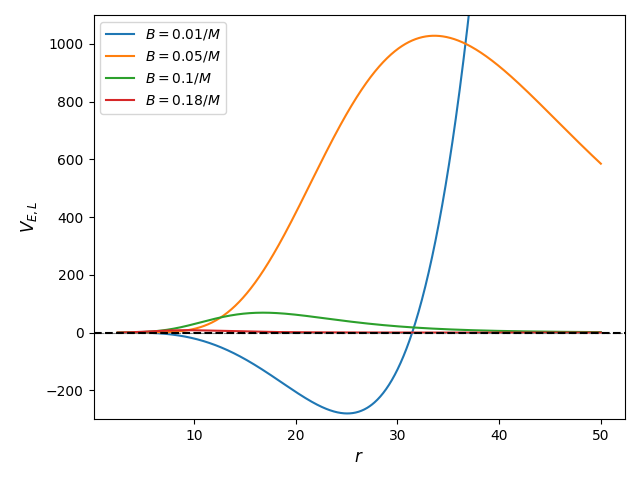}
    \caption{\label{fig:V(E,L)} { Effective potential $V_{E,L}$ from Eq. (\ref{eq: Timelike_Effective_Potential}) vs. $r$ for timelike geodesics for various $B$ values.} Here we give $r$ in units of $M$ and $V_{E,L}$ in units of $M^2$.  $E$ and $L$ were arbitrarily chosen as $1$ and $4M$.}
\end{figure}
For circular timelike { geodesics}, we get the equations for $E^2$ and $L^2$ in terms of $r$
\begin{equation}
     E^2=\frac{\Lambda\left(4Mr^2B^2-4r+8M+\frac{1}{4}B^4r^5+MB^4r^4\right)}{\left(1-\frac{2M}{r}\right)^{-1}(3B^2r^3-5MB^2r^2-4r+12M)},
    \label{eq:E(r)}
\end{equation}
\begin{equation}
     L^2=\frac{r^2}{\Lambda^2}\left[\frac{4Mr^2B^2-4r+8M+\frac{1}{4}B^4r^5+MB^4r^4}{ \Lambda(3B^2r^3-5MB^2r^2-4r+12M)}-1\right].
     \label{eq:L(r)}
\end{equation}
{ The condition that $L^2 > 0$ determines the radius values where circular timelike geodesics exist.}
In the Schwarzschild spacetime, circular timelike orbits exist for \(r>3M\). In the Ernst spacetime, however, one finds that \(L^2>0\) only in the range
\[
r_2 < r < r_1,
\]
where \(r_{1}\) and $r_2$ are the two positive roots of the cubic equation \eqref{eq:cubic_equation} indicating that circular timelike orbits can only exist between the two circular lightlike orbits in the equatorial plane. { In particular, this implies that circular timelike geodesics do not exist if $B>B_c$. Fig. \ref{fig:L^2} shows a plot of $L^2$ as a function of $r$ for a particular subcritical value of $B$.}
\begin{figure}[htbp]
    \centering
    \includegraphics[width=0.5\textwidth]{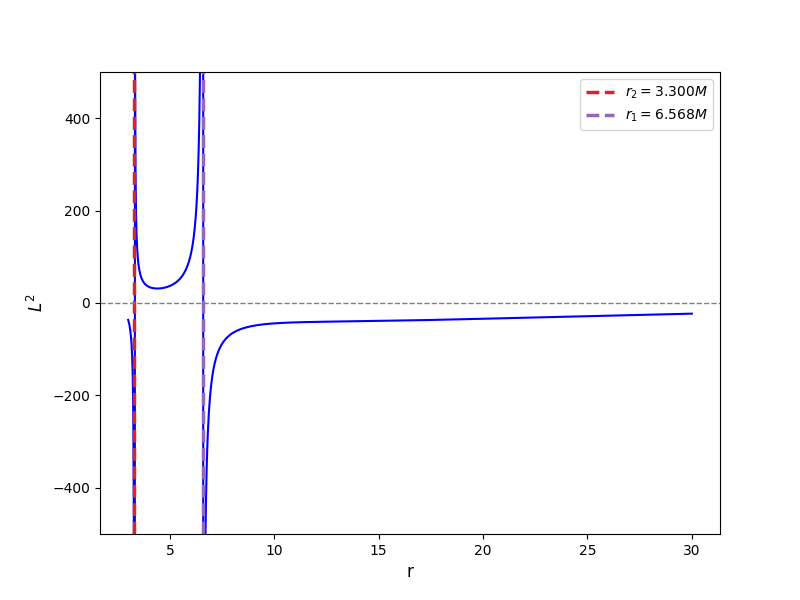}
    \caption{\label{fig:L^2}$L^2$ plotted { as a function of} $r$ for $B=0.15/M$. $L^2<0$ indicates no orbit. Here we give $r$ in units of $M$ and $L^2$ in units of $M^2$.}
\end{figure}

{ For determining which of the circular timelike geodesics are stable we have to consider the second derivative \(V''_{E,L}(r)\). One finds that \(V''_{E,L}(r)\) has exactly one zero, at a value \(r=r_{\mathrm{ISCO}}\), in the interval $r_2<r<r_1$ and is positive on the interval $r_{\mathrm{ISCO}}< r < r_1$, see Fig~\ref{fig:V(E,L)''} for an illustration. $r_\mathrm{ISCO}$ denotes the radius coordinate of the innermost stable circular orbit ($V_{E,L}'' (r_{\mathrm{ISCO}})=0$).} We identify the region of stable timelike circular orbits:
\begin{eqnarray}
    r_\mathrm{ISCO}<r<r_1 \, .
\end{eqnarray}
\begin{figure}[htbp]
    \centering
    \includegraphics[width=0.5\textwidth]{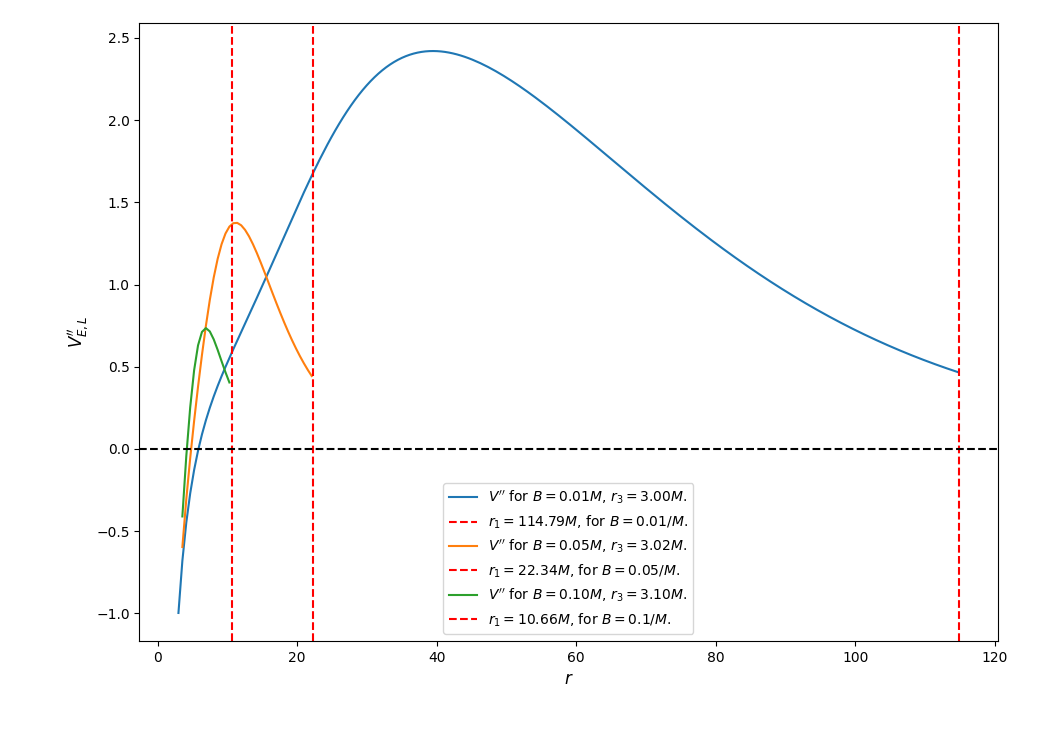}
    \caption{\label{fig:V(E,L)''}$V_{E,L}''$ vs. $r$ for various $B$ values. $V_{E,L}''=0$ corresponds to the ISCO. Here we give $r$ in units of $M$ and $V_{E,L}''$ in units of $M^2$. For each $r$, the energy $E$ and angular momentum $L$ of the circular timelike orbit were used, as given by Eqs.~(\ref{eq:E(r)}) and (\ref{eq:L(r)}), respectively.}
\end{figure}

%---------------------------------------------------------------------------------------
\section{\label{sec:Angular Deflection}Angular Deflection}
The deflection angle is usually defined for light rays that come in from infinity, go through a minimum radius value and then escape to infinity again. The deflection angle is determined by comparing the asymptote of the incoming ray to the asymptote of the outgoing ray. In Sec. \ref{sec:Geodesics in the Ernst Spacetime} we have seen that in the Ernst spacetime geodesics with $L \neq 0$ cannot reach infinity. For this reason, the standard definition of a deflection angle makes sense only for light rays with $L = 0$, i.e., for light rays that are confined to a meridional plane.

In a meridional plane, the null geodesics are identical to those in Schwarzschild, { recall Section \ref{subsec:Meridional Conformality}}. The trajectory equation for the meridional plane is given by:
\begin{eqnarray}
    \frac{d\tilde\phi}{dr}=\left[\frac{r^4}{\Lambda^8}\left(\frac{E^2}{{\tilde{L}}^2}-\frac{\Lambda^4\left(1-\frac{2M}{r}\right)}{r^2}\right)\right]^{-1/2}.
    \label{eq:dphi/dr}
\end{eqnarray}
For a light ray that goes through a minimum radius $r_m$ we have
\begin{eqnarray}
    \frac{E^2}{{ \tilde{L}}^2}=\frac{1}{r_m^2}-\frac{2M}{r_m^3}.
\end{eqnarray}
Using this expression in Eq.~(\ref{eq:dphi/dr}) and integrating over a light ray yields
\begin{equation}
    \Delta\tilde\phi_{\mathrm{mer}} \;=\; 2 \int_{r_m}^{\infty}
      \frac{r_m \, dr}{\sqrt{\bigl(1 - \tfrac{2M}{r_m}\bigr)\,r^4 
                            \;-\; r^2\,r_m^2 \;+\; 2M\,r\,r_m^2}},
\end{equation}
which defines the shift in the coordinate $\tilde\phi$ that the light ray experiences in the meridional plane. Unsurprisingly, this expression coincides exactly with the Schwarzschild result and, as in that case, the integral diverges in the limit $r_m \to 3M$.

{ Outside a meridional plane, the deflection  angle can be defined only for the case that the light source and the observer are at a finite distance. The resulting formula, which follows from integrating the orbit equation of a light ray over the (finite) section from the source to the observer, was given already several years ago by Konoplya \cite{Konoplya2007}.}

%------------------------------------------------------------------------------------------
{ 
\section{\label{sec:Shadow Angular Radii}ANGULAR RADII OF THE SHADOW}
}
{ As in the Ernst spacetime the lightlike geodesics are in general chaotic, the shadow cannot be calculated analytically. However, we will show in this section that it is possible to determine analytically the vertical angular diameter $\alpha$ of the shadow, and for an observer in the equatorial plane also the horizontal angular diameter $\beta$. For a numerical determination of the shadow in the Ernst spacetime we refer to Lima Junior et al. \cite{LimaJunior2022}.}
%We denote by $\alpha$ and $\beta$ as, respectively, the vertical (meridional) angular radius and the horizontal (equatorial) angular radius of the Schwarzschild-Melvin black hole shadow. 

For determining the shadow, it is crucial to specify the location of the light sources. One has to consider all initial directions of past-oriented light rays from the observer position; one assigns brightness to those light rays that do reach a light source and darkness to those that do not. In the case of an asymptotically flat spacetime, it is reasonable to assume that light sources are on a sphere of radius $r_S$ and to consider the limit $r_S \to \infty$. In the Ernst spacetime with $B \neq 0$, this is not reasonable because light rays with $L \neq 0$ never reach infinity, recall Section \ref{sec:Geodesics in the Ernst Spacetime}. So we assume that light sources are on a sphere of radius $r_S$, for a finite $r_S$.

%----------------------------------------------------------------------------------
\subsection{Vertical angular radius}
The vertical shadow radius $\alpha$ is determined { by past-oriented light rays from the observer position that spiral in a meridional plane towards the circular photon orbit at $r=3M$, so it is given by the same formula as in the Schwarzschild spacetime. The latter is known since many decades (although early papers on the subject did not use the word ``shadow''). It reads 
\begin{eqnarray}
    \sin^2\alpha=\frac{27M^2(r_O-2M)}{r_O^3},
\label{eq:alpha}
\end{eqnarray}
as was first given by Zel'dovich and Novikov \cite{ZeldovichNovikov1966} and almost simultaneously and independently by Synge \cite{Synge1966}}. Here $r_O$ denotes the radius coordinate of the observer who is assumed to be static. The $\vartheta$-coordinate of the observer is arbitrary. This equation is true for $r_S>3M$ and $2M<r_O<r_S$. $\alpha$ is a monotonically decreasing function of $r_O$, starting with the limiting value $\pi$ at $2M$ and going to the value $\pi /2$ at $r=3M$.
%----------------------------------------------------------------------------------
\subsection{Horizontal angular radius}
\begin{figure}[htbp]
    \centering
    \includegraphics[width=0.5\textwidth]{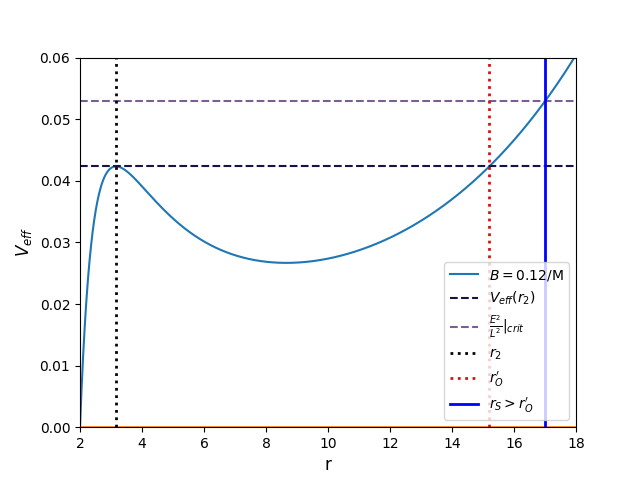}
    \caption{\label{fig:Effective_potential_B=0.12}Equatorial effective potential $V_\mathrm{eff}$ from Eq. (\ref{eq:Effective_potential}) vs. $r$ for $B=0.12/M$. Here we give $r$ in units of $M$ and $V_\mathrm{eff}$ in units of $1/M^2$.}
\end{figure}

\begin{figure}[htbp]
    \centering
    \includegraphics[width=0.5\textwidth]{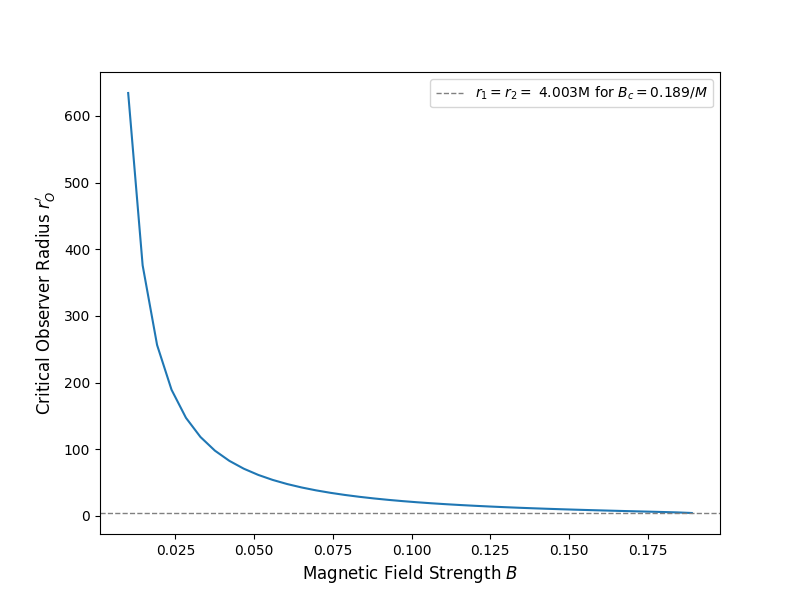}
    \caption{\label{fig:r_O'}Maximum observer radius $r_O'$ vs magnetic field strength $B$. Here we give $B$ in units of $1/M$ and $r_O'$ in units of $M$.}
\end{figure}

The horizontal shadow radius $\beta$ can be analytically determined only for an observer in the equatorial plane, $\vartheta _O = \pi /2$. We have to assume that the magnetic field is subcritical, $0 < B < B_c$, which guarantees the existence of a stable circular lightlike orbit at radius $r_1$ and an unstable one at radius $r_2$, where $r_2 < r_1$. We read from Fig. \ref{fig:Effective_potential_B=0.12} that there is a radius coordinate $r_O'$ with $r_2 < r_1< r_O'$, determined by the equation 
\begin{equation}
    V_{\mathrm{eff}} (r_2) = V_{\mathrm{eff}} (r_O') \, .
\end{equation}
If $B$ increases from $0$ to $B_c$, $r_O'$ decreases from infinity to the critical radius $r_c$, recall Eq. (\ref{eq:rc}).
The radius value $r_O'$ is marked by a red dotted line in Fig. \ref{fig:Effective_potential_B=0.12} and plotted in Fig. \ref{fig:r_O'} for different values of magnetic field strength $B$. 
The shadow construction depends on how the radius coordinate $r_O$ of the observer, whom we assume static again, and the radius coordinate $r_S$ of the light sources are related to $r_O'$. If we assume that $r_2<r_O<r_S<r_O'$, we read from Fig. \ref{fig:Effective_potential_B=0.12} that the boundary of the shadow is determined by light rays that asymptotically spiral towards the unstable photon circle at $r = r_2$. Correspondingly, the horizontal shadow radius is given by the equation
\begin{equation}
\sin^2\beta = \frac{\Lambda^4(r_O,\pi/2)(r_O - 2M)r_2^3}{\Lambda^4{(r_2,\pi/2)}(r_2 - 2M)r_O^3}.
\label{eq:beta}
\end{equation}
If $r_O$ varies from $r_2$ to $r_1$, the angle $\beta$ starts at the value $\pi/2$, decreases to a minimum and then increases again until reaching the value $\pi/2$ again at $r_O'$.

A further possibility arises when the light sources lie in the domain $r_1<r_S<r_O'$ such that the potential at \(r_S\) satisfies \(V_{\rm eff}(r_1)<V_{\rm eff}(r_S)<V_{\rm eff}(r_2)\). In this case, there is a second radius \(r_S'\in(r_2,r_1)\) with
\[
V_{\rm eff}(r_S')=V_{\rm eff}(r_S)\,.
\]
Consequently, an observer located at any radius \(r_O\in(r_S',r_S)\) finds that past-directed rays with
\[
\frac{E^2}{L^2}<V_{\rm eff}(r_S)
\]
cannot reach the light sources at \(r_S\); these rays remain trapped or fall inward and thus produce a dark patch on the observer’s sky. The boundary of that patch is given by the rays with the critical value \(E^2/L^2=V_{\rm eff}(r_S)\). Note that the inner matching radius \(r_S'\) exists only when \(V_{\rm eff}(r_S)\) lies strictly between the local minimum and local maximum of \(V_{\rm eff}\); if \(V_{\rm eff}(r_S)\le V_{\rm eff}(r_1)\) or \(V_{\rm eff}(r_S)\ge V_{\rm eff}(r_2)\) no such inner solution in \((r_2,r_1)\) exists.

By contrast, if we assume that the light sources are at a radius $r_S > r_O'$, as indicated by the blue solid line in Fig. \ref{fig:Effective_potential_B=0.12}, Eq. (\ref{eq:beta}) does not give the horizontal radius of the shadow.
In this case, we define a critical value of the effective potential
\begin{eqnarray*}
V_{\mathrm{eff}} (r_S)=\frac{E^2}{L^2}\bigg|_\mathrm{crit} \, ,
\end{eqnarray*}
corresponding to the point where the vertical line $r=r_S$ meets the plot of the potential. Hence, the light rays with $E^2/L^2<E^2/L^2\big|_\mathrm{crit}$ never reach $r_S$, so we associate darkness with these light rays, while light rays with $E^2/L^2>E^2/L^2\big|_\mathrm{crit}$ contribute to the bright region. The borderline case, which gives the boundary of the shadow, is given by the outwardly directed light rays with $E^2/L^2\big|_\mathrm{crit}$, so in this case $\beta$ is bigger than $\pi /2$ giving us the following equation for the horizontal component of the shadow
\begin{equation}
\sin^2\beta = \frac{\Lambda^4(r_O,\pi/2)(r_O - 2M)r_S^3}{\Lambda^4{(r_S,\pi/2)}(r_S - 2M)r_O^3}.
\label{eq:beta2}
\end{equation}

%---------------------------------------------------------------------------------------------------
\subsection{Geometry of the shadow}
{ The exact geometry of the shadow cannot be determined analytically. However, we may compare our analytical expressions for} the vertical and horizontal angular radii. We want to analytically prove that for an observer in the equatorial plane the shadow is oblate in the sense that $\beta > \alpha$. We first consider the case that $r_2 < r_O < r_S < r_O'$.
Then $\alpha$ and $\beta$ lie between $0$ and $\pi/2$, i.e., we have to prove that 
$\mathrm{sin} ^2 \beta  > \mathrm{sin} ^2 \alpha$. According to (\ref{eq:alpha}) and (\ref{eq:beta}),
this is equivalent to 
\begin{equation}
    \dfrac{\Big(1+B^2 r_O^2/4 \Big)^4}{\Big(1+B^2 r_2^2/4 \Big)^4}
    > 27 M^2 \, \dfrac{(r_2-2M)}{r_2^3} \, .
\end{equation}
This is indeed true, because the left-hand side is bigger than 1 while the right-hand side is smaller than 1. The first claim is obviously true because $r_O>r_2$. To verify the second claim, we observe that the right-hand side takes the value 1 for $r_2 = 3M$ and is a monotonically decreasing function of $r_2$ in the domain $r_2>3M$, as can be easily checked by calculating the derivative with respect to $r_2$. As $r_2>3M$, this implies that  the right-hand side is indeed smaller than 1.  In the case $r_O' < r_S$, for all observer positions between $r_2$ and $r_S$ the horizontal radius $\beta$ is bigger than $\pi /2$ while the vertical radius $\alpha$ is smaller than $\pi /2$, so the statement that $\beta > \alpha$ is true in this case as well.

\begin{figure}[htbp]
    \centering
    \subfigure{\includegraphics[width=0.46\textwidth]{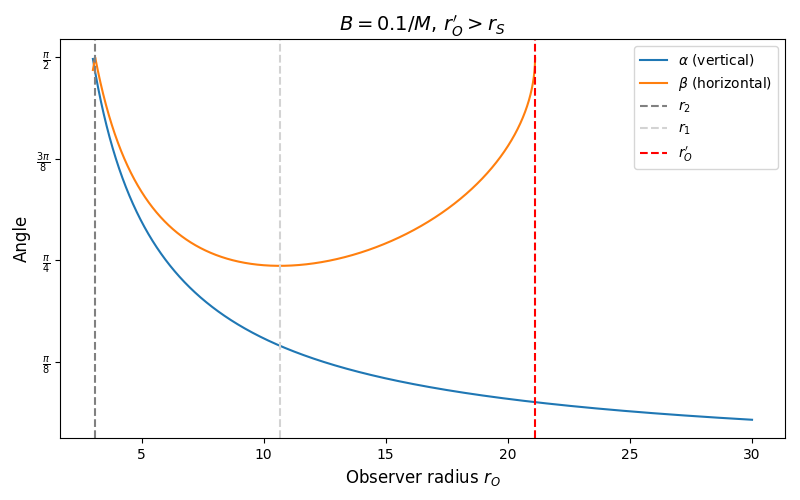}}
    \hfill
    \subfigure{\includegraphics[width=0.46\textwidth]{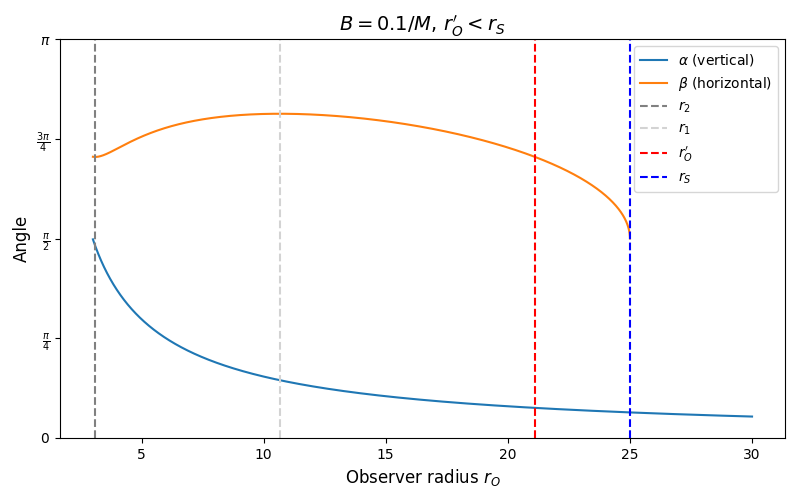}}
    \caption{\label{fig:Shadow_Radii} Shadow vertical and horizontal angular radii vs. observer distance $r_O$ in the equatorial plane. In the first plot, $r_S$ can be chosen anywhere $r_2<r_S<r_O'$. In the second plot, $r_S=25M$ was chosen above $r_O'$. Here we give $r_O$ in units of $M$ and the Angle in radians.}
\end{figure}

This confirms that for all $0 < B < B_{\mathrm{c}}$ and for all observers located in the equatorial plane at $r_2 < r_O < r_S$ the shadow exhibits an oblate shape. This analytical result is consistent with the numerical findings of Lima Junior et al. \cite{LimaJunior2022}. The two angular radii are plotted as a function of the observer radius $r_O$ for $B=0.1/M$ ($r_O'=21.12M$) for the two cases of $r_O'>r_S$ and $r_O'<r_S$ in Fig. \ref{fig:Shadow_Radii}.
%-----------------------------------------------------------------------------------------------------
\section{\label{sec:Photon Rings in Ernst Spacetime}PHOTON RINGS IN THE ERNST SPACETIME}

In this section, we consider an accretion disk in the equatorial plane of the Ernst spacetime. It is our goal to determine the visual appearance of this disk on the sky of an observer. For an arbitrary position of the observer, this problem cannot be solved analytically because the equation of motion for the light rays is chaotic. For a polar observer, i.e., for an observer on the axis of symmetry, however, an analytical solution is possible. The essential point is that for a polar observer, each light ray that reaches the observer stays in a meridional plane, so we can use the conformal equivalence to the Schwarzschild spacetime for analytically determining these light rays. {We stress, however, that this conformal equivalence does not imply that the photon rings in the Ernst spacetime are the same as the photon rings in the Schwarzschild spacetime. The reason is in the fact that the photon rings do not only depend on the paths of the relevant light rays; they also depend on the size of the luminous accretion disk. In particular, as identified in Sec. \ref{subsec: Circular Timelike Geodesics}, stable circular timelike geodesics in the Ernst spacetime with magnetic field $B$ exist only in the interval $r_{\rm ISCO}(B)<r<r_1(B)$, so an accretion disk composed of geodesic circular emitters is naturally restricted to $r_s\in(r_{\rm ISCO}(B),r_1(B))$ rather than to the Schwarzschild choice $r_s\in(6M,\infty)$.}  Thus, although we borrow the meridional (Schwarzschild) results for the polar-observer ray geometry, the Ernst spacetime exerts a secondary control via the permitted range of $r_s$ in the equatorial plane.

From each point of the accretion disk, there are infinitely many light rays to the observer, which are labelled by an integer $n=0,1,2,\dots$ which counts how often the light ray has crossed the axis before arriving at the observer. { We have mentioned already in the Introduction that} $n$ is known as the \emph{order} of the image. For an illustration of this situation, we refer to Fig. 1 in Bisnovatyi-Kogan and Tsupko \cite{BisnovatyiTsupko2022}. As before, we denote the azimuthal angle in the orbital plane of the light ray $\tilde{\phi}$, which should not be confused with the azimuthal angle in the equatorial plane, which is denoted $\phi$. Obviously, the azimuthal angle $\Delta \tilde{\phi}$ swept out by the light ray on its way from the light source to the observer is related to the order $n$ of the image by the simple relation
\begin{equation}
\Delta \tilde{\phi} = \Big( n+ \dfrac{1}{2} \Big) \, \pi \, ,
\label{eq:Polar_Observer}
\end{equation}
cf. again Fig. 1 in  Bisnovatyi-Kogan and Tsupko \cite{BisnovatyiTsupko2022}.

With the exception of the $n=0$ and $n=1$ cases, the considered light rays make more than one full turn around the center, so we can use the strong-deflection approximation in the version of Bozza and Scarpetta \cite{BozzaScarpetta2007} for calculating them. Because of the conformal equivalence, we can take the result from the literature where it had been done with the metric coefficients of the Schwarzschild spacetime. This gives us the following relation between the impact parameter $b = \tilde{L} /E$ and the azimuthal shift $\Delta \tilde{\phi}$ for a light ray that starts at radius coordinate $r_S$ and terminates at radius coordinate $r_O$:
\begin{equation}
    b=3 \sqrt{3} \left[1+F(r_s)F(r_O)\exp\left(-\Delta\tilde\phi \right)\right],
    \label{eq:b_phi}
\end{equation}
where 
\begin{equation}
    F(r) = \dfrac{
    6 \, \sqrt{6} \Big( 1 - \dfrac{3M}{r} \Big)
    }{
    2 + \dfrac{3M}{r} + \sqrt{3 + \dfrac{18M}{r}}
    } \, ,
\end{equation}
see Aratore et al. \cite{Aratore2024}, Eq. (17) and Sec. IIIA. 

Equating $\Delta \tilde{\phi}$ in this equation with the expression for $\Delta \tilde{\phi}$ from Eq. (\ref{eq:Polar_Observer}) gives us the impact parameter $b_n$ that corresponds to the $n$th order image, for $n \ge 2$,
    \begin{widetext}
    \begin{equation}
        b_n=3\sqrt3M\left[1+216\left(1-\frac{3M}{r_s}\right)\left(1-\frac{3M}{r_O}\right)\frac{e^{-\left(n+\frac{1}{2}\right)\pi}}{\left(2+\frac{3M}{r_s}+\sqrt{3+\frac{18M}{r_s}}\right)\left(2+\frac{3M}{r_O}+\sqrt{3+\frac{18M}{r_O}}\right)}\right].
    \label{eq:bn}
    \end{equation}
    \end{widetext}
{ Actually, this equation was given, already before the above-mentioned paper by Aratore \textit{et al.} \cite{Aratore2024}, by Bisnovatyi-Kogan and Tsupko \cite{BisnovatyiTsupko2022}, see Eqs. (15) and (16) in conjunction with Eq. (10) in their paper. The same formula can also be found in a paper by Tsupko \cite{Tsupko2022} as Eq. (24), but note that there is a misprint in the exponent.}

For an observer sufficiently far away from the center, the angle between the incoming ray and the vertical axis is $\approx b/r_O$. Then Eq. (\ref{eq:bn}) gives us the angular radius of the $n$th order image on the observer's sky of a luminous ring at radius $r_s$, for $n \ge 2$. For an extended luminous disk, we have to vary $r_s$ from the inner edge to the outer edge. { For each point of such a disk, there is an infinite sequence of images produced by the lensing effect of the black hole, which arrange in an infinite sequence of rings, called photon rings. In the next subsection we will discuss if and how the observation of these photon rings could be used for discriminating between a Schwarzschild and an Ernst black hole.}

%-------------------------------------------------------------------------------------------
\subsection{Gap parameter}
Broderick \textit{et al.} \cite{Broderick2022} proposed the relative size of a pair of photon rings to measure the spin of a Kerr black hole. Similarly, Wielgus \cite{Wielgus2021} considered the ratio of photon rings of order $n=1,2$ for various metrics and Eichhorn \textit{et al.} \cite{Eichhorn2023} suggested using the relative separation of photon rings as a probe of black hole spacetime.  { Building upon these ideas, Aratore \textit{et al.} \cite{Aratore2024} presented a systematic study of what they called the gap parameter, for spherically and static spacetimes. In the following we will demonstrate that this notion makes sense also in the Ernst spacetime, although the latter is not spherically symmetric.}

{ The gap parameter is defined as the relative separation between the radii of two consecutive photon rings normalized by the radius of the bigger ring:
\begin{equation}
    \Delta_n=\frac{b_n-b_{n+1}}{b_n}.
\end{equation}
Aratore \textit{et al.}} have obtained a range of variability of the gap parameter for various spacetimes by changing the value of $r_s$ from a chosen minimum value to infinity. The minimum radius in the case of Schwarzschild spacetime was chosen to be the Innermost Stable Circular Orbit (ISCO) given by $r_{\mathrm{ISCO}}=6M$ and the outer limit for $r_s$ was chosen as infinity. We focus on the first gap { parameter that can be calculated with the strong-deflection approximation,} $\Delta_2 = (b_2 - b_3)/b_2$, which measures the separation between the $n=2$ and $n=3$ rings.

We showed in Sec. \ref{subsec: Circular Timelike Geodesics} that stable timelike circular geodesics exist in the equatorial plane between the radii of the ISCO and the outer circular lightlike geodesics. If the luminous matter is in circular geodesic motion, we have to restrict $r_s$ correspondingly,
\[
r_s \;\in\; (\,r_\mathrm{ISCO},\,r_1\,).
\]
The assumption that the matter in an accretion disk moves on geodesics, i.e., that it behaves like a dust where pressure, viscosity, etc., are negligible, is usually a good approximation. We will make this assumption in the following. The fact that in the Ernst spacetime there is an upper bound for $r_s$ is the main difference in comparison to the Schwarzschild case.

%------------------------------------------------------------------------------------------------
\subsection{\label{Gap Parameter Intervals}Gap parameter intervals}
\begin{figure*}[htbp]
    \centering
    \includegraphics[width=1\textwidth]{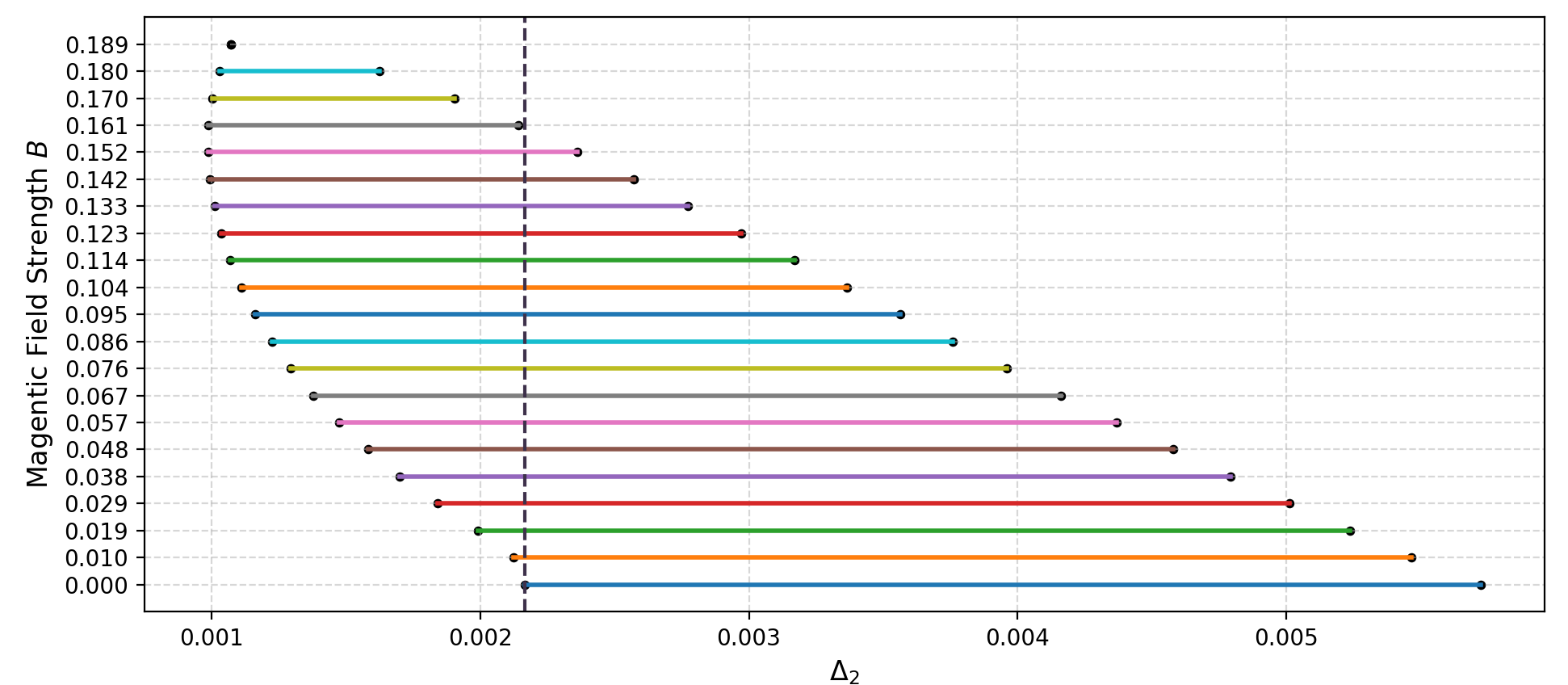}
    \caption{\label{fig:Gap_parameter}Intervals of the gap parameter $\Delta_2$, which is the relative separation between $n=2$ and $n= 3$ photon rings, for different values of the magnetic field strength $B$ in the Ernst spacetime.  Each line segment shows $\Delta_2$ as $r_s$ varies from the innermost stable orbit up to $r_1$ (outer photon orbit). The last line is the Schwarzschild case ($B=0$) while the top line is the critical case { ($B= B_c \approx 0.189/M$).} The vertical dashed line separates the Schwarzschild case from the { Ernst cases with $B>B_o \approx 0.160/M$.} $\Delta _2$ is dimensionless while $B$ is given in units of $1/M$.}
\end{figure*}

For a given $B$ value, we can evaluate the range of variability of the gap parameter for the Ernst spacetime for $r_s\in(r_\mathrm{ISCO},r_1)$. This is given by Fig. \ref{fig:Gap_parameter}. $B=0$ is the Schwarzschild case and here $r_s\in(6M,\infty)$ for $r_O=\infty$ to match with the results obtained in Aratore et al. \cite{Aratore2024}.

From Fig. \ref{fig:Gap_parameter} we can see that beyond a certain value { $B_o$ of the magnetic field strength $B$, the overlap between the range of gap parameter for the Ernst case and the Schwarzschild case is null. This limit value $B_o$ corresponds to the magnetic field strength of an Ernst spacetime in which the outer radius of the disk coincides with the ISCO of the Schwarzschild spacetime, $r_1=6M$. One determines this value numerically to be $B_o \approx 0.160/M$. It was emphasized by Aratore \textit{et al.} \cite{Aratore2024} that, to within the assumed approximations, the gap parameter is independent of the black hole mass and of the observer distance, In this sense, a measurement of $\Delta _2$ can unequivocally distinguish an Ernst spacetime with $B > B_o$ from a Schwarzschild spacetime. Of course, this conclusion is based on the -- not unrealistic -- hypothesis that the matter in the luminous disk moves on geodesics.}

{ We have restricted our discussion of photon rings and gap parameter in the Ernst spacetime to the case of a polar observer. For a non-polar observer the photon rings are, of course, non-circular and their exact shape and size cannot be determined analytically. However, even in that case it is possible to get some analytical results: One could use the equations derived by Aratore \textit{et al.} \cite{Aratore2024} for an observer at arbitrary inclination in the Schwarzschild spacetime for calculating the \emph{vertical} gap between photon rings. We will not work this out here.}

%-------------------------------------------------------------------------------
\section{\label{sec:Conclusion}Conclusion}
We { have presented} some analytical results on gravitational lensing in the Ernst spacetime, {thereby complementing} previous numerical studies. For these results, it was crucial that there are unstable circular lightlike geodesics in every meridional slice at $r = 3 M$ and that there are two circular lightlike geodesics, the inner one unstable and the outer one stable, in the equatorial plane outside the horizon for $0 < B < B_c$. We showed that, besides these circular lightlike geodesics, there are no other spherical lightlike geodesics in the Ernst spacetime.

The existence of the circular lightlike geodesics allowed us to analytically calculate the vertical angular radius of the shadow for an observer at arbitrary inclination and the horizontal angular radius of the shadow for an observer in the equatorial plane. The exact shape of the shadow cannot be determined analytically, but our results demonstrate that the shadow is always oblate for an observer in the equatorial plane, agreeing with the numerical findings of Lima Junior \textit{et al.} \cite{LimaJunior2022}. We showed how the oblateness of the shadow depends on the magnetic field. 

We also determined analytically, for a polar observer, the angular radii of the photon rings in the strong deflection approximation and discussed the dependence of the gap parameter $\Delta _2$ on the magnetic field. Once the $n = 2$ and $n = 3$ photon rings have been resolved in some { future} observations, this gives us the possibility to distinguish black holes described by the Ernst metric with a { sufficiently strong} magnetic field from Schwarzschild (and other) black holes.

{We believe that our analytical results on shadow oblateness and gaps between photon rings provide a concrete observational pathway to constraining the presence and approximate strength of magnetic fields in the neighborhood of supermassive black holes with near-future observations.}

\begin{acknowledgments}
{ Both authors are grateful to Oleg Tsupko for helpful discussions. Moreover, MHK acknowledges financial} support from the Erasmus Mundus Joint Master program in Astrophysics and Space Science.
\end{acknowledgments}
\appendix
\section{\label{app: Solutions of the Cubic Equation}Solutions of the cubic equation}
The three roots of the cubic equation (\ref{eq:cubic_equation}) are given by:\\
\begin{equation}
r_1= \frac{\left(X+ Y \right)^{\frac{1}{3}}}{9 B^2}- \frac{Z}{9 B^2 \left(X+Y \right)^{\frac{1}{3}}}+ \frac{5 M}{9}
\end{equation}
\begin{equation}
r_2 = \frac{- (1 + i \sqrt{3}) \left(X+Y\right)^{\frac{1}{3}}}{18 B^2} + \frac{(1 - i \sqrt{3})Z}{18 B^2 \left( X + Y \right)^{\frac{1}{3}}} + \frac{5 M}{9}
\end{equation}
\begin{equation}
 r_3= \frac{- (1 - i \sqrt{3}) \left(X+Y\right)^{\frac{1}{3}}}{18 B^2} + \frac{(1 + i \sqrt{3})Z}{18 B^2 \left( X + Y \right)^{\frac{1}{3}}} + \frac{5 M}{9}
\end{equation}
Where 
\begin{gather*}
    X=125 B^6 M^3 - 1188 B^4 M,\nonumber\\
    Y= 18 \sqrt{3} \sqrt{-375 B^{10} M^4 + 1352 B^8 M^2 - 48 B^6},\nonumber\\
    Z=-25 B^4 M^2 - 36 B^2.
\end{gather*}
\bibliography{Bibliography}

\end{document}